# Extended Eckart Theorem and New Variation Method for Excited States of Atoms

**Zhuang Xiong** [1,3 *], **Jie Zang**[1], **N.C. Bacalis**[2], **Qin Zhou**[3]

[1]Space Science and Technology Research Institute, Southeast University, Nanjing 210096, People's Republic of China

[2]Theoretical and Physical Chemistry Institute, National Hellenic Research Foundation, Vasileos Constantinou 48, GR-116 35 Athens, Greece

[3] School of Economics and Management, Southeast University, Nanjing 210096, People's Republic of China

* Email: zhuangx@seu.edu.cn

## Abstract

We extend the Eckart theorem, from the ground state to excited states, which introduces an energy augmentation to the variation criterion for excited states. It is shown that the energy of a very good excited state trial function can be slightly lower than the exact eigenvalue. Further, the energy calculated by the trial excited state wave function, which is the closest to the exact eigenstate through Gram–Schmidt orthonormalization to a ground state approximant, is lower than the exact eigenvalue as well. In order to avoid the variation restrictions inherent in the upper bound variation theory based on Hylleraas, Undheim, and McDonald [HUM] and Eckart Theorem, we have proposed a new variation functional $\Omega_n$ and proved that it has a local minimum at the eigenstates, which allows approaching the eigenstate unlimitedly by variation of the trial wave function. As an example, we calculated the energy and the radial expectation values of $^3S^{(e)}$ Helium atom by the new variation functional, and by HUM and Eckart theorem, respectively, for comparison. Our preliminary numerical results reveal that the energy of the calculated excited states $3\,^3S^{(e)}$ and $4\,^3S^{(e)}$ may be slightly lower than the exact eigenvalue (inaccessible by HUM theory) according to the General Eckart Theorem proved here, while the approximate wave function is better than HUM.

# 1. Introduction

The electrons around the nucleus in atoms are described by the Schrödinger wave equation. However, for many-particle systems Schrodinger equation cannot be solved exactly, from the time of establishment of quantum mechanics up to now; seeking a precise approximate solution of many-particle Schrodinger equation, for the ground state, and especially for the excited states, has been one of the most challenging research directions in physics [1-7]. Precise atomic structure calculations can not only explain the mechanism of electron correlation effects, but also provide the indispensable parameters of the relevant material science, such as plasma physics, astrophysics [8-10], which controlled nuclear fusion, etc. On the other hand, through the advances in vacuum technology, cryogenic technology, detection technology, etc, which provides more accurate experimental measurements, analyzing these high precision experimental results put higher requirements on theoretical calculations. Therefore, obtaining highly accurate wave functions, especially the wave functions of excited states, becomes an urgent need.

Through *ab-initio* theoretical methods, calculating the atomic structure is based on the variation principle [11]. The traditional standard method, utilizing variation theory to solve the Schrödinger equation for excited state wave functions, is based on the Hylleraas, Undheim, and McDonald [HUM] upper bound theorem [12-14]. In solving the secular equation in finite N-dimensional Hilbert space, the lowest root is the upper-limit of the ground state exact energy, while its higher roots are upper-limits of the corresponding upper excited states exact energies [15, 16] . All states obtained by this method are orthogonal to each other. However, as our previous work[17-21] showed, in N-dimensional Hilbert space this method has inherent restrictions, so that the 'quality' of the excited state wave function obtained by optimization will be lower than that of the ground state wave function.

In practical applications, it was found: On the one hand, if the ground state is optimized, the accuracy of the excited states, which are orthogonal to the ground state, will be reduced; if an excited state is optimized, then all orthogonal states lower than this excited state will loose accuracy. Thus, by either optimization in the ground state or in the excited state, it is impossible to get acceptable accuracy for the ground and the excited wave functions simultaneously, thereby, it will be difficult to achieve the high precision requirements of the experiment. On the other hand, if the calculation of the ground state and excited states are optimized by variation respectively [15], such as the State-Specific Theory (SST) developed by Nicolaides' group, based on approximated orthogonality, the energy of the ground state and the excited states can be obtained with good accuracy, but the orthogonality between the ground and excited state wave functions will be destroyed, and thus can not guarantee the accuracy of the approximate wave function. Particularly, this would render the physical calculation between the different states (e.g., optical transitions, the oscillator strength, etc.) unable to guarantee its reliability. This is why the spectral line positions of atoms and ions, obtained by many experiments and astronomical observations can be accurately theoretically explained, but their intensity distribution is difficult to be obtained with satisfactory theoretical description.

In order to overcome the intrinsic defects of HUM, we had proposed a new variation functional $\Omega_n$ [17], and based on this we developed a new variation algorithm [19, 22, 23] to get more accurate and reliable excited state wave functions. By calculating the helium (He) ground and first excited states of He $^3S^{(e)}$, we preliminarily numerically demonstrated the results of the General Eckart Theorem, and showed the intrinsic defects of HUM and the superiority of the new variation function.

According to the HUM theorem[12-14], for the $i$ th excited state, the usual practice is to optimize the basis set, especially when it contains nonlinear parameters, with respect to the $i$ th eigenvalue of the secular equation. But then any other root is not optimal; each must therefore be optimized individually: Additional imposition of orthogonality to the (correct) lower states (if possible), gives generally a better upper bound than the linear variation alone. In this attempt, the General Eckart

Theorem [20, 21] for excited states must be considered.

## 2. Theory and method

### 2.1 General Eckart Theorem

Let $\psi_1, \psi_2, \ldots, \psi_n$ be the exact eigenstates of the Hamiltonian $H$ (a complete orthonormal set) with energies $E[\psi_1] < E[\psi_2] < \cdots E[\psi_3] < \cdots$ , and let

$$\phi_n^{(e)} = \sum_{i=1}^{\infty} \langle \psi_i | \phi_n^{(e)} \rangle \psi_i, \qquad (1)$$

with

$$\sum_{i=1}^{\infty} |\langle \psi_i | \phi_n^{(e)} \rangle|^2 = 1 \qquad (2)$$

$\phi_n^{(e)}$ be the calculated normalized (n-1)th excited state approximation. Then

$$E[\phi_n^{(e)}] = \langle \phi_n^{(e)} | H | \phi_n^{(e)} \rangle = \sum_{i=1}^{n-1} |\langle \psi_i | \phi_n^{(e)} \rangle|^2 E[\psi_i] + \sum_{k=n}^{\infty} |\langle \psi_k | \phi_n^{(e)} \rangle|^2 E[\psi_k] \qquad (3)$$

Multiplying (2) by $E[\psi_n]$ and subtracting from (3), we obtain:

$$E[\psi_n] = E[\phi_n^{(e)}] + \delta_n^{(e)} - \varepsilon_n^{(e)} \qquad (4)$$

where both $\delta_n^{(e)}$ and $\varepsilon_n^{(e)}$ are positive (or zero if $\phi_n^{(e)} = \psi_n$):

$$\delta_n^{(e)} \equiv \sum_{i=1}^{n-1} |\langle \psi_i | \phi_n^{(e)} \rangle|^2 (E[\psi_n] - E[\psi_i]) \geq 0 \qquad (5)$$

$$\varepsilon_n^{(e)} \equiv \sum_{k=n+1}^{\infty} |\langle \psi_k | \phi_n^{(e)} \rangle|^2 (E[\psi_k] - E[\psi_n]) \geq 0 \qquad (6)$$

Since it is impossible to calculate $\varepsilon_n^{(e)}$, Equation (4) and (6) imply that

$$\varepsilon_n^{(e)} = (E[\phi_n^{(e)}] + \delta_n^{(e)}) - E[\psi_n] \geq 0 \qquad (7)$$

that is, the exact energy eigenvalue $E[\psi_n]$ is a lower bound of the calculated augmented energy:

$$E[\psi_n] \le (E[\phi_n^{(e)}] + \delta_n^{(e)}) = E[\phi_n^{(e)}] + \sum_{i=1}^{n-1} \left|\langle \psi_i | \phi_n^{(e)} \rangle\right|^2 (E[\psi_n] - E[\psi_i]) \qquad (8)$$

not of just the calculated expectation value $E[\phi_n^{(e)}]$. This is the General Eckart Theorem. For excited states, the two terms $\delta_n^{(e)}$ and $\varepsilon_n^{(e)}$ in (4) are competing and $E[\psi_n]$ may be either below or above $E[\phi_n^{(e)}]$, unless $\phi_n^{(e)} = \psi_n$ (which never happens). Therefore, any accidental coincidence, $E[\phi_n^{(e)}] = E[\psi_n]$ does not imply that $\phi_n^{(e)} = \psi_n$ if $\delta_n^{(e)} \ne 0$. This should be kept in the mind in any variation calculation of excited states. For the ground state, [(n=1), i.e., e=g], (7) reduces the usual Eckart upper bound theorem, since $\delta_n^{(g)} \equiv 0$.

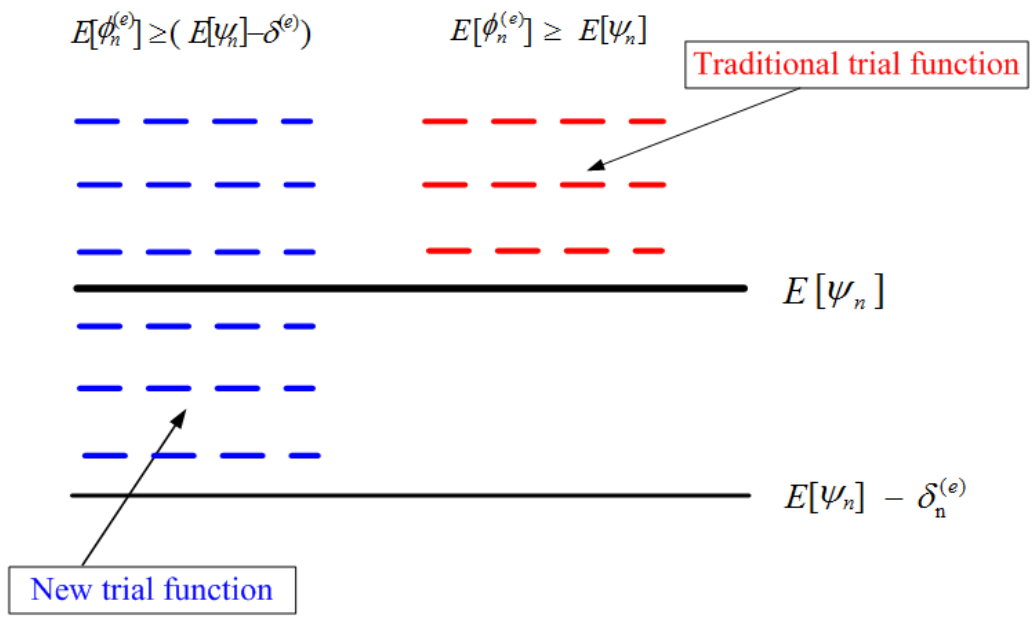

Figure 1. Schematic diagram of General Eckart Theorem (GET). The calculated excited energy by traditional variation method is the upper bound of its eigenvalue $E[\psi_n]$, whereas the calculated excited energy by exact trial function is the upper bound of $E[\psi_n] - \delta_n^{(e)}$, instead of $E[\psi_n]$.

As showed in the Figure 1, in practice, for excited states, the energy $E[\phi_n^{(e)}]$ of the calculated approximation does not have to lie above the exact $E[\psi_n]$; it may as well lie below, and between two approximate wave functions lying slightly above and slightly below the exact energy, the lower lying (even seemingly unacceptable:

$E[\phi_n^{(e)}] \leq E[\psi_n]$) would be more trustable if it had less augmented energy than the higher lying(!). All lower lying approximations should not be generically rejected; the one with the least augmented energy is the best approximation to $\psi_n$ (better than any higher lying).

Further, according to the HUM theorem, the higher roots of the secular equation tend to the excited state energies from above of the exact eigenvalue. But it should be observed that among all functions $\phi_1$, which are orthogonal to an available ground state approximant $\phi_0$, the Gram–Schmidt orthogonal to $\phi_0$ is $\phi_1^+$:

$$\phi_1^+ \equiv \frac{\psi_1 - \phi_0 \langle \psi_1 | \phi_0 \rangle}{\sqrt{1 - \langle \psi_1 | \phi_0 \rangle^2}} \tag{9}$$

which is the closest to the exact $\psi_1$ (the largest projection $|\langle \psi_1 | \phi_1 \rangle|^2$), lies energetically below the exact $E[\psi_1]$. Only if $E[\phi_0] < E[\psi_1]$:

$$E[\phi_1^+] = E[\psi_1] - \frac{(E[\psi_1] - E[\phi_0]) \langle \psi_1 | \phi_0 \rangle^2}{1 - \langle \psi_1 | \phi_0 \rangle^2} < E[\psi_1] \tag{10}$$

Therefore, the 2nd HUM root, $E[1^{HUM}]$, lying higher than $E[\psi_1]$, is necessarily not the closest to $\psi_1$ (while orthogonal to $\phi_0$).

Thus, in seeking a highly accurate approximate wave function of the excited state which is orthogonal to $\phi_0$, either by HUM or by Eckart Theorem, all cannot be achieved, simply because of the upper bound theory. The excited states cannot be obtained variationally by minimization of the energy, since they are saddle points in the Hamiltonian eigenfunction Hilbert space. So the traditional variation method must be modified. Therefore, a new variation functional $\Omega_n$ which has local minimum at the excited state has been proposed, and the approximate wave function obtained can be infinitely close to the corresponding eigenstate through searching for minimum of $\Omega_n$.

## 2.2 The variation function $\Omega_n$

The presented functional $\Omega_n$ is

$$\Omega_n \equiv E[\phi_n] + 2[\sum_{i<n} \frac{(E[\phi_n]\langle\phi_i|\phi_n\rangle - \langle\phi_i|H|\phi_n\rangle)^2}{E[\phi_n] - E[\phi_i]}] / [1 - \sum_{i<n}\langle\phi_i|\phi_n\rangle^2] \quad (11)$$

where $\phi_i$ are approximants of $\psi_i$.

We have proved[17] that for a non-degenerate Hamiltonian of bound eigenstates of a specific type of symmetry, $\psi_1, \psi_2 \cdots$ with eigenenergies $E[\psi_1] < \cdots < E[\psi_n]$, the functional $\Omega_n$ of Eq. (11) has a local minimum at $\phi_n = \psi_n$, when $\phi_i \approx \psi_i$, while $E[\phi_n]$ has a saddle point there.

In fact, $\Omega_n$ has a local minimum at $|\phi_n\rangle = |\psi_n\rangle$, even if $|\psi_i\rangle, (i<n)$ are not accurate approximants of the exact eigenfunctions. (Ref. [15] p.282) This made the practical calculations possible. So $|\phi_n\rangle$ is allowed to approach $\psi_n$ at will, just like $|0^{HUM}\rangle$ approach $|\psi_0\rangle$:

$$|\langle\psi_n|\phi_n\rangle|^2 \leq 1 \quad (12)$$

## 3. Computing and Results Analysis

In order to do the numerical study, in this work, we extend our Generalized Laguerre Type Orbital (GLTO) atomic study-package by using the new variation functional $\Omega_n$ method. The GLTO atomic study-package [19,22,23] is recently developed with economical and comprehensive, generalized Laguerre type atomic orbital (GLTO), which provide conciseness, clear physical interpretation and near equivalent accuracy with numerical multi-configuration self-consistent field(MCHF), one of the most accurate numerical methods for atomic CI calculations.

The computing program based on functional $\Omega_n$ method is developed as the flow chart shown on Figure 2, where the rough low-lying excited state $\phi_1, \phi_2 \cdots$ can be simply calculated from our GLTO atomic package. This flow chart is similar with our previous work [22, 23] reported on reference.

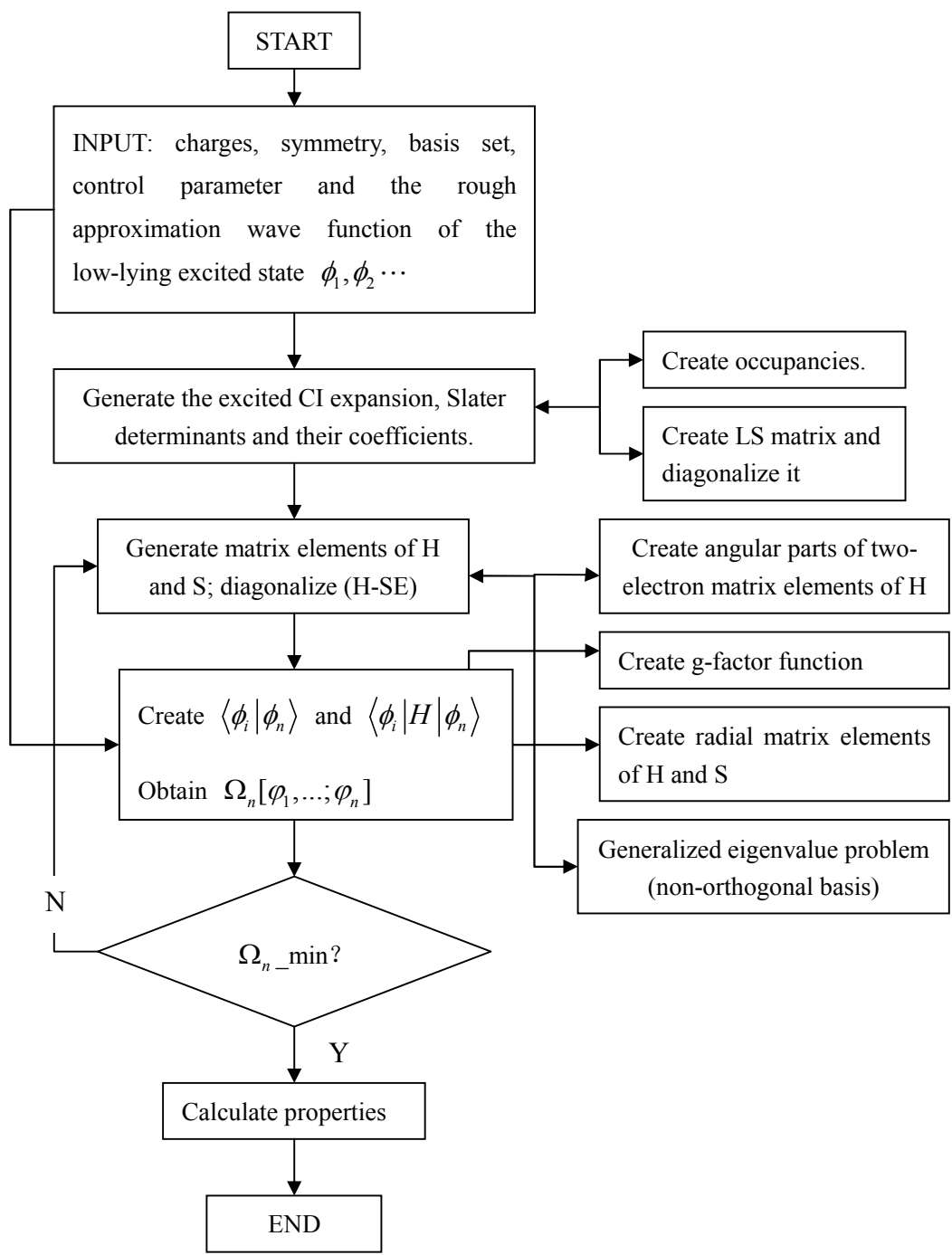

Figure.2. Flow chart of the calculating program

For a many-electron atom its eigenstates cannot be calculated exactly, so it is difficult to demonstrate the quality of calculated atomic wave functions directly. However, calculating their radial expectation values $\langle \phi_n | r^m | \phi_n \rangle, (m = -1, 1, 2)$ can be used to indicate the accuracy of calculated atomic trial wave functions. [24]

By using our developed GLTO atomic calculating package, which includes the codes developed by HUM theory, Eckart Theorem and the new variation function $\Omega_n$ respectively, we study helium (He) atom's wave functions and the energy values of the ground state and the first excited state of $^3S^{(e)}$, and calculate the relevant radial expectation values.

In order to demonstrate the quality of calculated atomic wave functions by the different methods, we use the same GLTO orbitals up to 5f, configurations (Cs) and Slater determinants (SDs) in our calculations by HUM, and $\Omega_n$ functional method respectively. We use, as standard reference, the most reliable calculated atomic data obtained by Chen M K (1994) [24] by B-splines method using more than 100 B-splines functions. We demonstrate the quality of the atomic wave function with different methods by calculating their relative error respectively.

Table 1. The energy values and radial expectation values of the lowest and of the first two excited state of He $^3S^{(e)}$, in atomic units.

|  |  | Variation for $\Omega$ | HUM ($2^3S$) | Chen M K | Relative | error | Exact |
|---|---|---|---|---|---|---|---|
|  |  | $E[\phi_n]$ | H | C | $\left\|\dfrac{E[\phi_n]-C}{C}\right\|$ | $\left\|\dfrac{H-C}{C}\right\|$ | $E[\psi_n]$ |
| $2^3S$ | $E_1$ | -2.1752135 | -2.1752135 | -2.1752288 | $7.033\times10^{-6}$ | $7.033\times10^{-6}$ | -2.1753598 |
|  | $<1/r>$ | 1.15465 | 1.15465 | 1.154664 | $3.464\times10^{-6}$ | $3.464\times10^{6}$ |  |
|  | $<r>$ | 2.55057 | 2.55057 | 2.550468 | $3.999\times10^{-5}$ | $3.999\times10^{-6}$ |  |
|  | $<r^2>$ | 11.46337 | 11.46337 | 11.46438 | $8.810\times10^{-5}$ | $8.810\times10^{-5}$ |  |
| $3^3S$ | $E_2$ | -2.0693402 | -2.0562612 | -2.0686888 | $3.149\times10^{-4}$ | $6.007\times10^{-3}$ | -2.0688059 |
|  | $<1/r>$ | 1.05683 | 1.07955 | 1.063674 | $6.434\times10^{-3}$ | $1.493\times10^{-2}$ |  |
|  | $<r>$ | 6.01157 | 4.60073 | 5.855982 | $2.656\times10^{-2}$ | $3.157\times10^{-1}$ |  |
|  | $<r^2>$ | 71.16894 | 40.48480 | 68.70871 | $3.580\times10^{-2}$ | $4.101\times10^{-1}$ |  |
| $4^3S$ | $E_3$ | -2.0366305 | -1.8968475 | -2.0365120 | $5.818\times10^{-5}$ | $6.858\times10^{-2}$ | -2.0366209 |
|  | $<1/r>$ | 1.03128 | 1.15230 | 1.034570 | $3.180\times10^{-3}$ | $1.138\times10^{-1}$ |  |
|  | $<r>$ | 10.77838 | 3.42043 | 10.66123 | $1.098\times10^{-2}$ | $6.792\times10^{-1}$ |  |
|  | $<r^2>$ | 239.54449 | 26.02633 | 238.580 | $4.042\times10^{-3}$ | $8.909\times10^{-1}$ |  |

In table 1, the first column shows variation results obtained, optimizing the lowest root of the $\Omega_n$ secular equation, following Eckart Theorem. We calculated the wave function $3^3S$ at the minimum of $\Omega_2$, by using 13 GLTOs (1s, 3s, 4s, 5s, 2p, 3p, 4p, 5p, 3d, 4d, 5d, 4f, 5f), full CI with 16 (Cs), 46 Slater determinants (SDs) and the first order approximation wave function $\phi_1$ is calculated with 2GLTOs (1s,2s), 1 Cs, 1 SD, while its energy $E[\phi_1]=-2.1718648$; We calculated the wave function $4^3S$ at the minimum of $\Omega_3$, by using 12 GLTOs (1s, 4s, 5s, 2p, 3p, 4p, 5p, 3d, 4d, 5d, 4f, 5f), full CI with 13 (Cs), 43 Slater determinants (SDs) and the first order approximation wave function $\phi_1$ is calculated with 2GLTOs (1s,2s), 1 Cs, 1 SD, while its energy $E[\phi_1]=-2.1718648$, and the first order approximation wave function $\phi_2$ is calculated with 2GLTOs (1s,3s), 1 Cs, 1 SD, while its energy $E[\phi_2]=-2.0685536$); we calculated the wave function $2^3S$ at the minimum of $\Omega_1$, by using 14 GLTOs (1s, 2s, 3s, 4s, 5s, 3p, 4p, 5p, 3d, 4d, 5d, 4f, 5f), full CI with 20 Cs,

50 SDs.

The HUM($2^3$S) column results are calculated, according to the theory of HUM, by optimizing the first root of the $E_n$ secular equation as ground state $2^3$S and then the second root and the third root are regarded as an approximant of the state $3^3$S and $4^3$S respectively, the HUM calculation is carried out with 14 GLTOs (1s, 2s, 3s, 4s, 5s, 3p, 4p, 5p, 3d, 4d, 5d, 4f, 5f), full CI with 20 Cs, 50 SDs as well; the third column shows reference values. In the fourth and fifth columns the relative errors of the two different methods compared with reference $C$ are given. In the last column the latest MCHF standard values (See http://nlte.nist.gov/MCHF/index.html, National Institute of Standards and Technology (NIST) MCHF Database are given.

From this table, for the first excited state $3^3$S using the same CI expansions, one can see that the data obtained by the new variation functional, which is free from the inherent restrictions of the HUM [17], is more precise than the data obtained by HUM either in the energy or in the radial average value, even the former has an order of magnitude higher accuracy. Besides, the new variation functional energy value (-2.0693402 a.u.) is lower than the MCHF standard value (-2.0688059 a.u.), while, of course, the $\Omega_2$ value itself is a little higher (-2.0679501a.u.), and the relative error of the energy value is smaller than that of HUM. Both facts imply that a trial wave function for excited state may be approximated in the circumstance that the energy value is less than the eigenvalue.

Similarly for the second excited state $4^3$S using the same CI expansions, it is shown that the wave function obtained by the new variation functional, which is free from the inherent restrictions of the HUM [17], is more precise than the data obtained by HUM, and the new variation functional energy value (-2.0366305 a.u.) is lower than the MCHF standard value (-2.0366209 a.u.), while the $\Omega_3$ value itself is a little higher (-2.0363067 a.u.). It is demonstrated, as the General Eckart theorem proved, that for the excited state, the lower lying wave function (even seemingly unacceptable by HUM theory: $E[\phi_n^{(e)}] \leq E[\psi_n]$) would be more trustable if it had less augmented energy than the higher lying.

Using our wave functions $\phi_i$, which are calculated by $\Omega_n$, let $\psi_i \approx \phi_i$, then our estimated $\delta_n^{(e)}$-augmentation values are $\delta_1^{(e)} = \delta^{(g)} = 0$ a.u. for the $2^3S$ state, $\delta_2^{(e)} = 6.819 \times 10^{-4}$ a.u. for the $3^3S$ state, and $\delta_3^{(e)} = 2.0528 \times 10^{-4}$ a.u. for the $4^3S$ state, which are calculated by Eqs.(5), one can easily exam that these data consistent with our theoretical expectations Eqs.(8). i.e in table 1 $E[\psi_n] \leq (E[\phi_n^{(e)}] + \delta_n^{(e)})$ (n = 1, 2, 3) These are numerical examination for General Eckart Theorem.

Finally, for the lowest state $2^3S$, i.e., the ground state of symmetry $^3S$, we notice that the calculation results based on function $\Omega_n$ and based on HUM are the same, as we expected, simply because, for the ground state, the functional $\Omega_n$ reduces to $E_n$, and the General Eckart Theorem reduces to Eckart Theorem.(cf. Equations (11) and (7))

## 4. Conclusions

As a conclusion, we can see that for the excited state, as General Eckart Theorem proved, the lower lying approximate wave function (even seemingly unacceptable by HUM theory: $E[\phi_n^{(e)}] \leq E[\psi_n]$) would be more accurate than the higher lying. It should be kept in mind that in any variation calculation of excited states, unlike the ground state, the energy $E[\phi_n^{(e)}]$ of the calculated approximation do not have to lie above the exact $E[\psi_n]$, all lower lying approximations should not be generically rejected, the one with the least augmented energy is the best approximation. Therefore, by seeking the minimum of the higher root energy $E[\phi_n^{(e)}]$ according to the traditional variation method we can not get a highly accurate approximate wave function for the excited state, whereas by seeking the minimum of the functional $\Omega_n$, we can do it.

Using traditional variation method to calculate the atomic structure through *ab-initio* theory is based on the HUM theorem. We can notice that from the theoretical analysis and this preliminary numerical calculation, for the excited state, the calculation is poor in accuracy although it had a strict theoretical foundation, while the method of the functional $\Omega_n$, not only had rigorous theoretical proof, but also is more trustable in truncated space, and, of course, in principle, the most accurate wave function of the excited states can be achieved by improving the variation wave function through $\Omega_n$ by adding more CI terms. It is the good hope to demonstrate the superiority of the functional $\Omega_n$ theory more lucidly by improving our computing program so as to run on supercomputing facilities.

## Acknowledgements

This work was partially supported by the Key Project of National Social Science Foundation of China (Grant No.15AJL004) and Polynano-Kripis 447963 / GSRT, Greece